\documentclass[runningheads]{llncs}

\usepackage{eccv}

\usepackage{eccvabbrv}

\usepackage{graphicx}
\usepackage{booktabs}

\usepackage[accsupp]{axessibility}  

\usepackage{hyperref}

\usepackage{orcidlink}
\begin{document}

\title{Neural Implicit Representations for 3D Synthetic Aperture Radar Imaging} 
\titlerunning{Neural Implicit Representations for 3D SAR}
\author{Nithin Sugavanam\orcidlink{0000-0001-9910-2258} \and
Emre Ertin\orcidlink{0000-0001-7815-0728} 
}

\authorrunning{N. Sugavanam and E. Ertin}

\institute{Ohio State University, Columbus, OH  43210, USA 
\email{\{sugavanam.3,ertin\}@osu.edu}}

\maketitle

\begin{abstract}
Synthetic aperture radar (SAR) is a tomographic sensor that measures 2D slices of the 3D spatial Fourier transform of the scene. In many operational scenarios, the measured set of 2D slices does not fill the 3D space in the Fourier domain, resulting in significant artifacts in the reconstructed imagery. Traditionally, simple priors, such as sparsity in the image domain, are used to regularize the inverse problem. In this paper, we review our recent work that achieves state-of-the-art results in 3D SAR imaging employing neural structures to model the surface scattering that dominates SAR returns. These neural structures encode the surface of the objects in the form of a signed distance function learned from the sparse scattering data. Since estimating a smooth surface from a sparse and noisy point cloud is an ill-posed problem, we regularize the surface estimation by sampling points from the implicit surface representation during the training step. We demonstrate the model's ability to represent target scattering using measured and simulated data from single vehicles and a larger scene with a large number of vehicles. We conclude with future research directions calling for methods to learn complex-valued neural representations to enable synthesizing new collections from the volumetric neural implicit representation. 
\end{abstract}

\section{Introduction}
Synthetic aperture radar (SAR) is a tomographic sensor that measures 2D slices out of the 3D Spatial Fourier transform of the scene.  Traditional 3D reconstruction techniques involve aggregating and indexing phase history data in the spatial Fourier domain and applying an inverse 3D Fourier Transform to the data as shown in references~\cite{ferrara2009enhancement,austin2010sparse}. Obtaining high-resolution imagery requires the data collected to be densely distributed in both azimuth and elevation angle, which is often not satisfied in operational scenarios. For instance, the sampling in the elevation dimension is sparse and non-uniform in the GOTCHA dataset~\cite{ertin2007gotcha,austin2009sparse}. To cope with the sparsely sampled data in elevation, regularized inversion methods~\cite{sugavanam2016recovery} that combine a non-uniform fast Fourier transform method to model the forward operator with regularization priors for the scene that promote a structured solution such as sparsity~\cite{sugavanam2017interrupted}, limited persistence in the viewing angle domain~\cite{sugavanam2017limited,sugavanam2018approximating,sugavanam2019compressing}, vertical structures~\cite{sugavanam2022models}. These regularization-based approaches promote dominant scattering mechanisms~\cite{sugavanam2022high,agarwal2020sparse} and produce sparse point clouds in the spatial domain.

In addition, scattering from multiple internal reflections can appear as scattering centers away from the physical surface of the object. Moreover, non-uniform sampling in elevation introduces ambiguities in the height direction, leading to aliased copies of the object. In this paper, we review our recent work that achieves state-of-the-art results in 3D SAR imaging employing neural structures to model the surface scattering that dominates SAR returns. These neural structures encode the surface of the objects in the form of a signed distance function learned from the sparse scattering data. Since estimating a smooth surface from a sparse and noisy point cloud is an ill-posed problem, we regularize the surface estimation by sampling points from the implicit surface representation during the training step. We demonstrate the model's ability to represent target scattering using measured and simulated data from single vehicles and larger scenes containing hundreds of objects. We conclude with future research directions calling for methods to learn complex-valued neural representations to  synthesize tomographic projections from previously unseen viewpoints/apertures.

\section{Related Work}

\subsection{Classical 3D SAR Imaging}
A common approach for $3D$ imaging is to formulate the inversion of the Fourier operator as an inverse problem imposing regularization to enforce sparsity in the spatial domain and promote correlation of scattering coefficients in the neighboring  sub-apertures~\cite{scarnati2021three}. Reference~\cite{ferrara2009enhancement} solves the recovery problem over individual sub-apertures and non-coherently integrates the result to obtain a wide-angle $3D$ representation of the object. The backscattered response of the object is also modeled as a superposition of the backscattered response of $3D$  canonical scattering mechanisms such as dihedral, trihedral, plate, cylinder, and top-hat~\cite{jackson2012synthetic}. The scattering behavior of these canonical reflectors has been derived as a function of the size of the scattering mechanism using the predictions from the Geometric theory of diffraction~\cite{Canonical3D_SAR_Rigling}. References~\cite{anisotropySAR_cetin_2006,sugavanam2017limited,sugavanam2022models} jointly model the sparsity in scattering center locations and the persistence of scattering coefficients in the azimuth domain. Alternatively, the imaging problem has been posed as an interferometric imaging problem using measurements obtained from multiple baselines~\cite{austin2009sparse} and~\cite{ertin2008multibaseline}. The 3D non-uniform Fourier transform is approximated by a set of 2D non-uniform Fourier transforms for each baseline for range and cross-range estimation and a 1D non-uniform Fourier transform for height estimation. The effect of the persistence of the target on the ambiguities in the point-spread function is presented in Reference~\cite{Potter_3dSAR_2007}. The uncertainty in the localization of scattering centers is dictated by the persistence, which leads to ambiguities in the target localizations along the vertical direction projected along the viewing angle of elevation.

\subsection{Neural Implicit Representations }
Meanwhile, in computer vision,  implicit neural representations, like Neural Radiance Fields (NeRF) and its derivatives~\cite{mildenhall2020nerf,wang2021nerf}, have become popular volume rendering methods, boasting strong performance even when dealing with highly intricate objects. This volume rendering approach uses focal-plane camera geometry to sample multiple points along rays and perform composition of the colors of the sampled points to link the 3D model to 2D views. The traditional discrete representations of objects, scene geometry, and appearance using meshes and voxel grids scale poorly with the scene's size. Recent developments utilize continuous functions parameterized by deep neural structures. These coordinate-based Deep-nets are trained to map the low-dimensional spatial coordinates to output a representation of shape or density for each spatial location. Coordinate-based Deep-Nets have been used to represent  images~\cite{tancik2020fourier}, volume density~\cite{wang2021nerf}, occupancy\cite{mescheder2019occupancy}, and signed distance~\cite{park2019deepsdf}.
The signed distance function (SDF) is given by
 \begin{align}\label{eq:SDF_def}
     SDF(\boldsymbol{p}) = \begin{cases}
     & 0, \boldsymbol{p} \in \Omega \\
     & +s, \boldsymbol{p} \in \Omega^{+}\\
     &-s, \boldsymbol{p} \in \Omega^{-},
     \end{cases}
 \end{align}
 where $s>0$ and $\Omega$ represents the object boundary, $\Omega^{+}$ represents the region outside the object and $\Omega^{-}$ represents the region inside the object.

Typically a Fourier feature layer is used as the input layer in the coordinate-based Deep-Nets that operate on the spatial coordinates. It has been shown that the Fourier feature mapping can be used to overcome the spectral bias of coordinate-based Deep-Nets towards low frequencies by allowing them to learn much higher frequency details in the geometry. Reference~\cite{tancik2020fourier} showed that a random Fourier feature mapping with an appropriately chosen scale could dramatically improve the performance of coordinate-based Deep-Nets across many low-dimensional tasks in computer vision.

\section{Denoising sparse point clouds using surface priors}
\label{sec:problemDescription}
We present a method to recover the point-cloud representation of the object from SAR phase history data. The measurements from the SAR are collected over azimuth angles $[\theta_{1,e},\cdots,\theta_{N_P,e}]$ and elevation angles $[\phi_{1,e},\cdots,\phi_{N_p,e}]$, with elevation passes $e=1,\cdots,N_{el}$ and frequency points $[f_1,\cdots,f_{N_F}]$. The phase history data are denoted by $\mathbf{Y}_e=[\mathbf{Y_{1,e}} \mathbf{Y_{2,e}} \cdots \mathbf{Y_{N_P,e}}]$ where $\mathbf{Y}=[\mathbf{Y}_1;\cdots;\mathbf{Y}_{N_{el}}] \in \mathbb{C}^{N_F \times N_P \times N_{el}}$. We model the object as a collection of dominant scattering centers, and each scattering center is isotropic over a sub-aperture. We assume the pulses collected over different viewing angles are grouped in $N_s$ sub-apertures over $N_{el}$ elevation passes. The backscattered signal in a sub-aperture $m$ is defined as 
\begin{align}
    &\bar{Y}(f_i,\theta_{j,e},\phi_{j,e};\bar{\theta}_m,\bar{\phi}_m) = \sum_{k=1}^K s^m_k \exp \left(-\frac{j4\pi f_i}{c} \right. \nonumber \\  &\left. \left( x_k \cos\phi_{j,e}\cos\theta_{j,e} +y_k \sin\theta_{j,e}\cos\phi_{j,e} + z_k\sin\phi_{j,e}\right)\right), \nonumber \\
    & \qquad \qquad + n(f_i,\theta_j,\phi_k) 
\end{align}
where $i=1,\cdots,N_F$, $j=1,\cdots,N_P$,$e=1,\cdots,N_{el}$, $m=1,\cdots,N_S$, $\bar{\theta}_m$ is the mean azimuth angle in sub-aperture $m$ and $\bar{\phi}_m$ is the mean elevation angle in sub-aperture $m$ . 
The K-sparse scattering centers are located at spatial coordinate $\boldsymbol{p}_k=[x_k, y_k, z_k]$ with sub-aperture dependent scattering coefficients $s^m_k \in \mathbb{C}$. The region of interest is discretized into $N_x \times N_y \times N_z$ voxels, and the resolution of these grid points is chosen according to the range resolution to get cube-like voxels. The measurement noise and modeling mismatches are combined in the complex-valued noise term $n(f_i,\theta_{j,e},\phi_{j,e})$. The sub-aperture measurements can be expressed as
\begin{align}
\boldsymbol{Y}(\bar{\theta}_m,\bar{\phi}_m) = \mathcal{F}^m_{3D}(\boldsymbol{S}_m) +\boldsymbol{N},
\end{align}
where $\boldsymbol{Y}(\bar{\theta}_m,\bar{\phi}_m) \in \mathbb{C}^{N_F \times N_S} \times N_{el}$ is the sub-aperture phase history data, $\mathcal{F}^m_{3D}(.)$ is the non-uniform Fourier transform~\cite{greengard2004accelerating} that maps the spatial coordinates to the K-space measurement space, and $\boldsymbol{S}_m \in \mathbb{C}^{N_x \times N_y \times N_z}$ are the complex-valued scattering coefficients for each spatial location in the region of interest.

Given the measurements, the inverse problem of recovering the scattering coefficients is solved using the regularized least squares method. The sparsity-promoting prior is used to enhance the dominant scattering centers and suppress the side lobes generated from the point spread function of the measurement operator. The regularized least squares problem is given by
\begin{align}
\min_{\mathbf{S}_m} \lVert \mathbf{S}_m\rVert_1 \qquad \textbf{Subject to } \lVert \boldsymbol{Y}(\bar{\theta}_m) - \mathcal{F}^m_{3D}(\boldsymbol{S}_m)\rVert^2 \leq \sigma^2,
\end{align}
where $\sigma^2$ is the modeling error energy. The recovered scattering centers have a sparse voxel representation for the sub-aperture $m$. The results from all the sub-apertures are combined non-coherently to get a joint-sparse representation given by
\begin{align}
\mathbf{S} = \sum_{m=1}^{N_s} \lvert  \mathbf{S}_m \rvert.
\end{align}
The resulting quantity $S$ consists of the scattering coefficients of the point-cloud representation of the object in a regular voxel space. We denote each point in the voxel space with scattering coefficients above a minimum threshold belonging to the set $\mathcal{\boldsymbol{P}}$. Specifically, we denote the point-cloud $\mathcal{\boldsymbol{P}}$ representing the object as 
\begin{align}
\mathcal{\boldsymbol{P}}= \left \{\boldsymbol{p}_i = [x,y,z] \textbf{ if } \lvert S([\boldsymbol{p}_i]) \rvert \geq \tau \right \}.
\end{align}
Next, we also find the viewing direction that corresponds to the maximum response of each point across different sub-apertures. We denote this set as $\mathcal{\boldsymbol{V}}$ and each $\boldsymbol{v}_i$ is given by
\begin{align}
m^{*} =& \text{argmax}_m \lvert \boldsymbol{S_m}(\boldsymbol{p}_i) \rvert \nonumber \\
\boldsymbol{v}_i =& [\cos(\bar{\theta}_{m^{*}})\cos(\bar{\phi}_{m^{*}}) ; \sin(\bar{\theta}_{m^{*}})\cos(\bar{\phi}_{m^{*}}) ; \sin(\bar{\phi}_{m^{*}} )]
\end{align} 

The initial estimate of the normal $\boldsymbol{n}_i \in \boldsymbol{\mathcal{N}_P}$ for each point $\boldsymbol{p}_i \in \mathcal{\boldsymbol{P}}$ is computed using a local principal component analysis. For each point, we find the nearest neighbors with a distance of at most $0.3m$. Next, we fit a plane using that subset of points, where the normal direction of the plane is assigned as the normal direction for the given point. Suppose the point $\boldsymbol{p}_i$ has fewer than $3$ neighbors in that neighborhood. We select the normal direction based on the sub-aperture with the maximum scattering coefficient at location $\boldsymbol{p}_i$. We utilize the set $\boldsymbol{V}$ and assign $\boldsymbol{n}_i=\boldsymbol{v}_i$. 

We utilize the signed distance function defined in Eq.~\eqref{eq:SDF_def} to represent the object and learn a coordinate-based MLP to predict the zero-level set of the SDF. Since the SAR point cloud is sparse, we use the concept of iso-points developed in Reference~\cite{yifan2021iso}. The iso-point set $\boldsymbol{\mathcal{Q}}_{iso}$ are the points sampled from the zero-level set of the neural network such that
\begin{align}
\boldsymbol{\mathcal{Q}}_{iso} = \left \{ \boldsymbol{q} : f(\boldsymbol{q},\Psi) = 0 \right \}
\end{align}
The sampling process to obtain dense samples on the iso-surface involves the following steps: projection, uniform sampling, and up-sampling. The projection operation is performed on a randomly sampled point $\boldsymbol{q}$ in the neighborhood of set $\mathcal{\boldsymbol{P}}$ to project it back on the surface represented by a zero-level set of the neural network $f(\boldsymbol{q};\Psi)$ using Newton's iterations. Given a point $\boldsymbol{q}_{k-1}$, the $k-th$ update step is given by 
\begin{align}\label{eq:newtonUpdateProject}
\boldsymbol{q}_{k}=\boldsymbol{q}_{k-1} -\boldsymbol{\pi}\left( \frac{ \boldsymbol{J}^T(\boldsymbol{q}_{k-1})}{\lVert \boldsymbol{J}(\boldsymbol{q}_{k-1}) \rVert^2} f(\boldsymbol{q}_{k-1})\right)
\end{align}
such that $J(\boldsymbol{q}_i)$ is the Jacobian of the network parameters with respect to the spatial coordinates $\boldsymbol{q}_i$. 
\begin{align}
\boldsymbol{\pi(\boldsymbol{x})}=\frac{\boldsymbol{x}}{\lVert \boldsymbol{x}\rVert} \min (\lVert \boldsymbol{x} \rVert,\tau_0)
\end{align}
$\boldsymbol{\pi(\boldsymbol{x})}$ is the clipping operation to avoid the possible noisy update step obtained from the non-smooth neural implicit surface representation, and $\tau_0$ is the maximum preset threshold computed based on the bounding box size around the region of interest. Newton's update steps are repeated until   $f(\boldsymbol{q}_k) \leq 1e-4$. 

This projection operator still produces non-uniform samples from the surface and can lead to regions with no samples. We obtain uniform samples on the surface by moving the points away from the high-density region such that
\begin{align}
\boldsymbol{q} = \boldsymbol{q}-\alpha \sum_{\boldsymbol{q}_k \in \mathcal{B}_{\epsilon}(\boldsymbol{q})} w(\boldsymbol{q}_i,\boldsymbol{q})\frac{\boldsymbol{q}_i-\boldsymbol{q}}{\lVert \boldsymbol{q}_i- \boldsymbol{q}\rVert },
\end{align}
where $\mathcal{B}_{\epsilon}(\boldsymbol{q})$ is the set of points in the ball of radius  $\epsilon$ around the point $\boldsymbol{q}$. We additionally define the weighting factor $w(\boldsymbol{q}_i,\boldsymbol{q})$ as
\begin{align}\label{eq:weightPointsDistance}
w(\boldsymbol{q}_i,\boldsymbol{q}) = \exp\left(-\frac{\lVert \boldsymbol{q}_i- \boldsymbol{q} \rVert^2}{\sigma_p^2}   \right),
\end{align}
such that distant points have reduced influence, where $\sigma_p$ is the density bandwidth chosen per the scene size. Following the uniform sampling step to move samples away from the high-density region, the resulting set of points is projected on the zero-level set of the SDF using the iterations in Eq.~\eqref{eq:newtonUpdateProject}.

Finally, a dense point cloud is created for a desired density using a simplified version of the edge-aware resampling (EAR) technique~\cite{huang2013edge}. The points are resampled such that these points are pushed away from the edges to avoid discontinuity in the normal direction while penalizing the formation of clusters. The normal for each point in the iso-point set is computed using the normalized Jacobian given by $\boldsymbol{n}_i  = \frac{J(\boldsymbol{q}_i)}{\rVert J(\boldsymbol{q}_i) \lVert}$. The estimated normal direction is used to update the iso-point cloud to ensure the points are away from edges by
\begin{align}
\Delta\boldsymbol{q}_{edge}= \frac{\sum_{\boldsymbol{q}_k \in \mathcal{B}_{\epsilon}(\boldsymbol{q})} \phi(\boldsymbol{n}_i,\boldsymbol{q}_i-\boldsymbol{q})\left(\boldsymbol{q}_i-\boldsymbol{q}\right)}{\sum_{\boldsymbol{q}_k \in \mathcal{B}_{\epsilon}(\boldsymbol{q})} \phi(\boldsymbol{n}_i,\boldsymbol{q}_i-\boldsymbol{q})},
\end{align}
where $\phi(\boldsymbol{n}_i,\boldsymbol{q}_i-\boldsymbol{q})=\exp\left(-\frac{\boldsymbol{n}_i^T(\boldsymbol{q}_i-\boldsymbol{q})}{\sigma_p^2}\right)$ is the anisotropic projection weight such that the normal computed for all the neighboring points $\boldsymbol{q}_i$ is used to prefer points that lie on the same plane, enforce smoothness, and penalize edges. The following term penalizes the grouping of points and utilizes the weight function defined in Eq.~\eqref{eq:weightPointsDistance}.  
\begin{align}
\Delta\boldsymbol{q}_{repulsion}= 0.5\frac{\sum_{\boldsymbol{q}_k \in \mathcal{B}_{\epsilon}(\boldsymbol{q})} w(\boldsymbol{q}_i,\boldsymbol{q})\left(\boldsymbol{q}_i-\boldsymbol{q}\right)}{\sum_{\boldsymbol{q}_k \in \mathcal{B}_{\epsilon}(\boldsymbol{q})} w(\boldsymbol{q}_i,\boldsymbol{q})}.
\end{align}
Points that are not well-separated are given higher weights such that they distribute uniformly on the surface. The update step for each point is given by
\begin{align}
\boldsymbol{q} = \boldsymbol{q} - \boldsymbol{\pi}(\Delta\boldsymbol{q}_{repulsion}) - \boldsymbol{\pi}(\Delta\boldsymbol{q}_{edge}).
\end{align}
Upsampling is performed by adding more samples to low-density point regions. A priority score to assess the regions is defined for each point $\boldsymbol{q}$ as
\begin{align}
P(\boldsymbol{q}) &= \max_{\boldsymbol{q}_i \in B_{\epsilon}(\boldsymbol{q})} \lVert \boldsymbol{q}_i - \boldsymbol{q} \rVert \\
\boldsymbol{q}^{*} &= \text{argmax}_{\boldsymbol{q}_i \in B_{\epsilon}(\boldsymbol{q})} \lVert \boldsymbol{q}_i - \boldsymbol{q} \rVert.
\end{align}
The $\boldsymbol{q}^{*}$ falls in the region of highest priority; therefore, we next find the highest priority neighbor of $\boldsymbol{q}^{*}$ by  $\boldsymbol{q}_i^{*} = \text{argmax}_{\boldsymbol{q}_i \in B_{\epsilon}(\boldsymbol{q}^{*})} \lVert \boldsymbol{q}_i - \boldsymbol{q}^{*} \rVert$. A new point is inserted using the following asymmetric rule to avoid duplicate copies  
\begin{align}
\boldsymbol{q}_{new} = \frac{\boldsymbol{q}_i^{*} + 2\boldsymbol{q}^{*}}{3}.
\end{align}
After the upsampling step, the generated points are again projected back on the SDF using Newton's iterations in Eq.~\eqref{eq:newtonUpdateProject}.
\subsection{Solution Framework}
\label{sec:solution}
In this section, we present the network architecture to predict the signed distance function and discuss the loss function to learn the network parameters. 
\subsubsection{Network architecture}
   \begin{figure} [ht]
   \begin{center}
   \begin{tabular}{c} 
   \includegraphics[height=4cm]{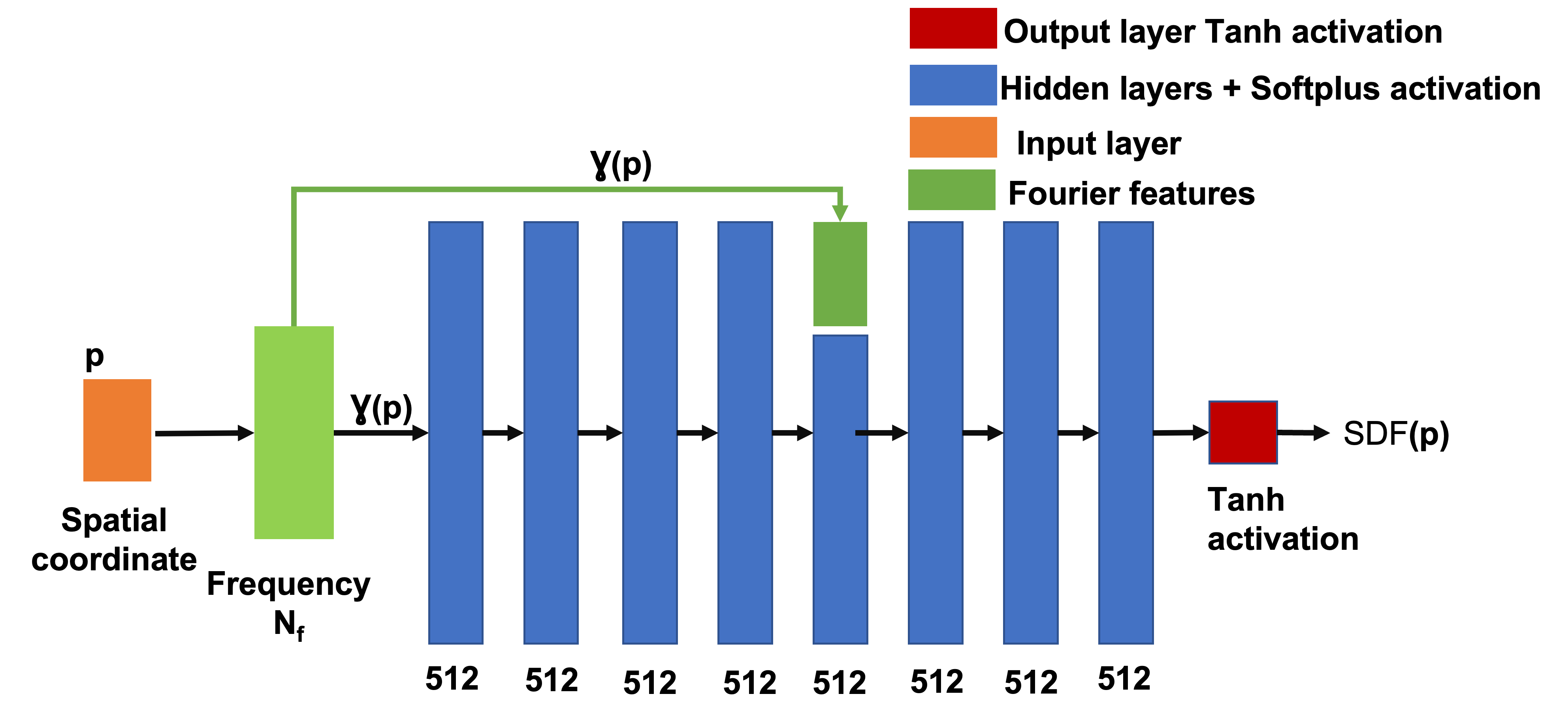}
   \end{tabular}
   \end{center}
   \caption[example] 
   { \label{fig:networkArchitecture} 
The network architecture consists of the input layer Fourier layer and $8$ linear layers with a softplus activation function and an output layer with a $\tanh$ activation function to predict the signed distance function.}
   \end{figure} 
We utilize the network architecture shown in Figure~\ref{fig:networkArchitecture} proposed in~\cite{park2019deepsdf} and~\cite{yifan2021iso} to estimate the signed distance function. The input Fourier feature layers map the spatial coordinates to $N_F$ frequencies. In the results section, we verify the effect of varying $N_F$ with the scene complexity. The hidden layers have a latent-feature size of $512$ with a soft-plus activation function. The transformed input layer is also fed in layer $4$. The output layer consists of a linear map and the $\tanh$ activation function to predict the SDF. Finally, the weights and biases of the network is initialized with standard Gaussian variables. 
\subsubsection{Loss function}
The detected point clouds from SAR measurements are sparse and noisy, since these are aliased copies due to non-uniform elevation sampling. These point clouds are usually concentrated near edges and other dominant scattering mechanisms. We aim to reconstruct the 3D object by enforcing the surface prior and denoising the point cloud by estimating the network parameters to learn a 3D representation of the object of interest. The iso-points estimated from the implicit representation serve as a consistent and smooth approximation of the object that is refined during the training process. This regularization ensures the training procedure does not overfit to noise. Since the iso-points are updated throughout the training process, we hypothesize the network learns the high-frequency signals governed by the underlying geometry of the object. We define three sets of point-clouds $\boldsymbol{\mathcal{Q}}_{iso}=\left\{ \boldsymbol{q} : f(\boldsymbol{q}; \Psi) = 0\right\}$ is the point-cloud consisting of iso-points, $\boldsymbol{\mathcal{P}}$ is the point-cloud consisting of the training set and $\boldsymbol{\mathcal{Q}}_{b}$ is the point-cloud consisting of the points sampled from the region of interest using a uniform probability distribution.
Since the iso-points are uniformly distributed on the surface, these points can be used to augment the supervision in undersampled areas by enforcing SDF to be $0$ on the iso-points set. The $0$ condition on the SDF is implemented through a sparsity promoting $\ell_1$ regularization in Eqs.~\eqref{eq:isosdf} and~\eqref{eq:onsdf}. The SDF for non-surface points is enforced to be non-zero by utilizing the exponential term from Eq,~\eqref{eq:offsdf}. The Eikonal regularizer in Eq.~\eqref{eq:eikonal} enforces the output of the network to mirror the properties of a SDF. 
\begin{align}
\label{eq:isosdf} L_{isoSDF} &= \frac{1}{\lvert \boldsymbol{\mathcal{Q}}_{iso} \rvert} \sum_{\boldsymbol{q} \in \boldsymbol{\mathcal{Q}}_iso}\lvert f(\boldsymbol{q}) \rvert \\
\label{eq:onsdf} L_{on} &=\frac{1}{\lvert \boldsymbol{\mathcal{P}}\rvert} \sum_{\boldsymbol{p} \in \boldsymbol{\mathcal{P}}}\lvert f(\boldsymbol{p}) \rvert \\
\label{eq:offsdf} L_{off} &=\frac{1}{\lvert \boldsymbol{\mathcal{Q}}_{b}\rvert} \sum_{\boldsymbol{q} \in \boldsymbol{\mathcal{Q}}_{b}} \exp \left( -\alpha \lvert f(\boldsymbol{p}) \rvert \right) \\
\label{eq:eikonal}L_{Eik} &=\frac{1}{\lvert \boldsymbol{\mathcal{Q}}_{background} \bigcup \boldsymbol{\mathcal{Q}}_{iso} \rvert} \sum_{\boldsymbol{q} \in \boldsymbol{\mathcal{Q}}_{b} \bigcup \boldsymbol{\mathcal{Q}}_{iso}}  \lvert 1 - \lVert \boldsymbol{J}^T(\boldsymbol{q}) \rVert \rvert 
\end{align}
We compute the iso-points' normals using a principal component analysis (PCA). The consistency between the Jacobian of the model and the normal estimated using a local neighborhood is evaluated using the cosine similarity as shown below
\begin{align}
L_{isoNormal}&=\frac{1}{\lvert \boldsymbol{\mathcal{Q}}_iso \rvert} \sum_{\boldsymbol{q} \in \boldsymbol{\mathcal{Q}}_iso} \left( 1-\lvert  SC\left( \boldsymbol{J}^T(\boldsymbol{q})         , \boldsymbol{n}_{PCA}(\boldsymbol{q})\right)\rvert \right) \\
L_{Normal}&=\frac{1}{\lvert \boldsymbol{\mathcal{P}}\rvert} \sum_{\boldsymbol{p} \in \boldsymbol{\mathcal{P}}} \left( 1-\lvert  SC\left( \boldsymbol{J}^T(\boldsymbol{p}), \boldsymbol{n}_{PCA}(\boldsymbol{p})\right)\rvert \right),
\end{align}
where the cosine similarity $SC(\boldsymbol{a},\boldsymbol{b}) =\frac{ \boldsymbol{a}
. \boldsymbol{b}}{\lVert\boldsymbol{a} \rVert  \lVert \boldsymbol{b} \rVert} $.  The optimization objective is comprised of six parts:
\begin{align}
&L =  \lambda_{isoSDF} L_{isoSDF} + \lambda_{isoNormal} L_{isoNormal} + \lambda_{eik} L_{eik}\nonumber \\  &  \lambda_{onSDF} L_{onSDF} + \lambda_{normal} L_{normal} + \lambda_{offSDF} L_{offSDF} .
\end{align}

\section{Results}
\label{sec:results}
We evaluate our algorithm on the Civilian Vehicle Data Domes dataset presented in~\cite{dungan2010civilian} and the GOTCHA parking lot measured datasets~\cite{ertin2007gotcha,austin2009sparse} using the entire circular aperture. We utilize $8$ elevation passes of a SAR sensor with 640 MHz bandwidth corresponding to roughly  0.25 cm resolution. The point-cloud and surface normals are recovered using regularized SAR inversion methods described in~\cite{austin2010sparse} . The point-cloud and the estimated normal vectors are utilized to train the network to estimate the SDF. 
The surface reconstruction by denoising the point-clouds is obtained using presented method. We qualitatively evaluate the performance of the surface reconstruction by varying the number of frequencies of the random Fourier feature layer at the input and the effect of utilizing iso-points in the training procedure.

\begin{figure} [h!]
   \begin{center}
   \begin{tabular}{c} 
   \includegraphics[width=0.8\linewidth]{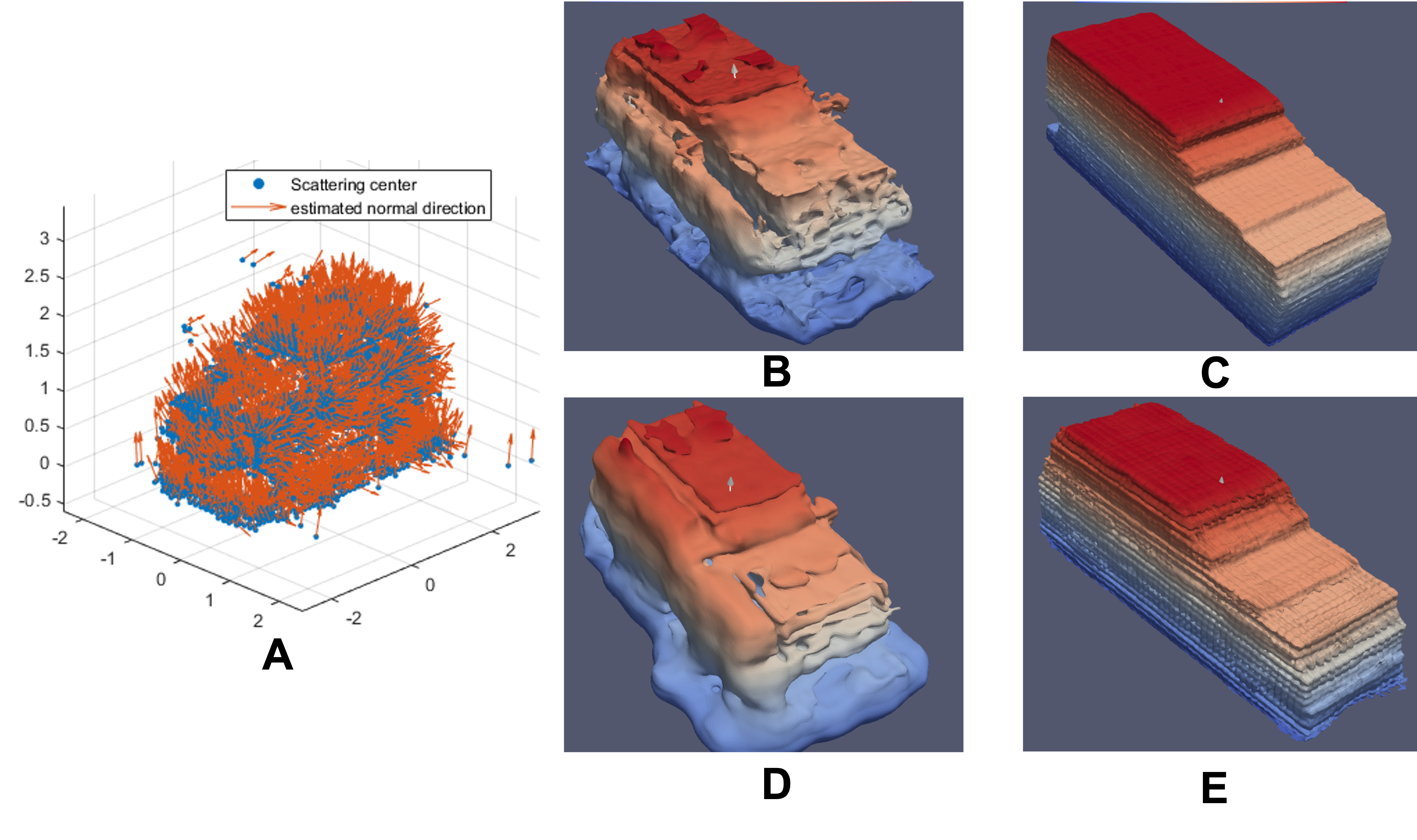}
   \end{tabular}
   \end{center}
   \caption
   { \label{fig:Jeep93_SDF} 
Figure. A refers to the detected point-cloud and the estimated normal vectors from SAR phase-history measurements for Jeep 93 Vehicle using 8 elevation passes. Figure. B and Figure. C illustrate the recovered mesh using the SDF network after training when iso-points are not generated from the network to regularize the network optimization for the number of frequencies as $6$ and $9$ in the input layer, respectively. Figure. D and Figure. E illustrate the recovered mesh using the SDF network after training when iso-points are  generated from the network to regularize the network optimization for the number of frequencies as $6$ and $9$ in the input layer, respectively. }
   \end{figure}

Figure~\ref{fig:Jeep93_SDF} illustrates the surface representation estimated by the network for a Jeep vehicle. We note that the points that are not close to the surface that arise from multi-path and non-uniform frequency domain sampling are treated as outliers and do not influence the surface estimation.  We vary the number of frequencies $N_f$ in the input layer. We consider $N_f = \{6,9\}$ for evaluation. We notice that as $N_f$ increases the edges are clearly defined and the flat surfaces are also modeled. We also note that fine details such as rear-view mirrors are missing for $N_f =9$ compared with $N_f=6$. Additionally, there are more artifacts and holes in the surface representation for the case of $N_f=6$ compared to the case of $N_f=9$. 
We also evaluate the benefit of utilizing iso-points in the training procedure. The iso-points regularize the network training procedure and help in estimating the surface representations with more details and avoids artifacts as shown in Figures~\ref{fig:Jeep93_SDF}.  

\begin{figure*} [th]
   \begin{center}
   \begin{tabular}{c} 
   \includegraphics[width=0.8\linewidth]{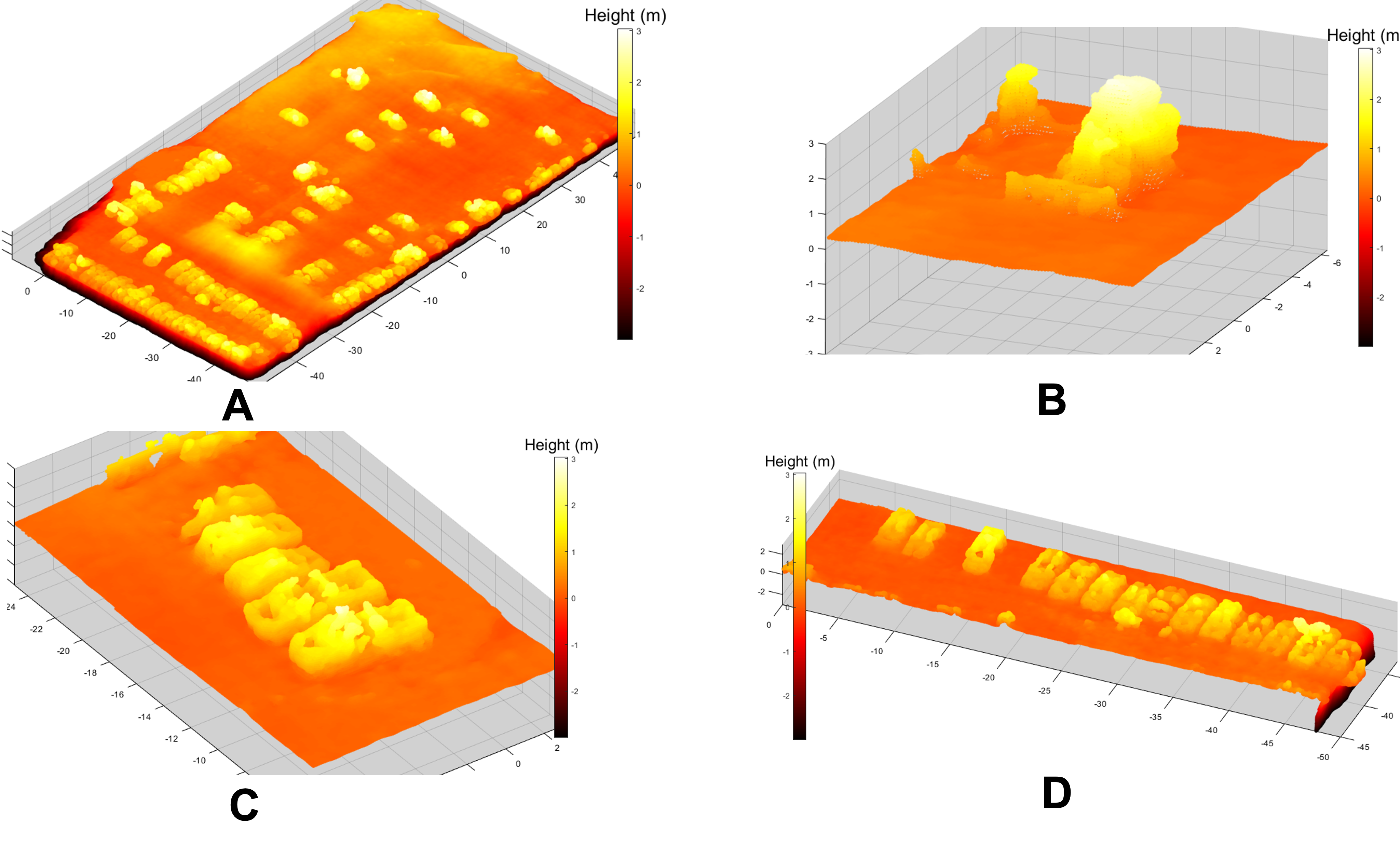}
   \end{tabular}
   \end{center}
   \caption[example] 
   { \label{fig:GotchaResults} 
Implicit surface recovery from the GOTCHA parking lot dataset. Figure A shows the entire parking lot, Figure B shows the backhoe from the dataset, and Figures C and D show the surface approximation of a set of cars in the parking lot.}
   \end{figure*}

Figure~\ref{fig:GotchaResults} shows the surface recovered from the GOTCHA parking lot dataset. We utilize the entire $360$ degree circular aperture to recover the point-cloud representation, creating sub-apertures with an azimuth span of $5$ degrees. The number of random Fourier features in our reconstruction is set to $N_F=9$. We observe that the proposed network architecture can successfully model the entire parking lot and capture the geometric details of individual vehicles. Further visualizations are provided in the supplementary materials.
\section{Future Research Directions: Complex valued Neural Implicit Representations}
The reconstruction framework reviewed  in this paper allows denoising of the point clouds produced by the SAR sensor. However, complex amplitudes of the scatterers are lost along the way and therefore does not allow synthesis of novel views of the constructed objects. This objective can be achieved by extending the  NeRF volume modeling strategy in two distinct ways to account for differences between Electro-optic (EO) and SAR phenomenology. First,  the color map function has to be replaced with  a view-dependent complex valued  reflectance map  with the view encoded by azimuth and elevation angles of the SAR platform. Second,  the simple ray tracing based focal plane projections of the EO model  has to be replaced with orthographic projection of 3D scatterers of the object onto the 2D slant measurement plane. We  note that the SAR projection operator is not a simple orthographic summation of the complex reflectivities; instead, contributions  are weighted with the phase terms proportional to their height from the slant plane, complicating the forward measurement model. This complex valued volumetric neural implicit representation can then be used to synthesize  tomographic projections from previously unseen viewpoints/apertures.  

\bibliographystyle{splncs}
\bibliography{report}
\end{document}